\documentclass[a4paper,12pt]{article}       
\usepackage{latexsym}

\begin{document}

\title{The local and global geometrical aspects of the twin paradox in static spacetimes: 
II. Reissner--Nordstr\"{o}m and ultrastatic metrics} 
     
\author{Leszek M. SOKO\L{}OWSKI and Zdzis\l{}aw A. GOLDA \\
Astronomical Observatory, Jagiellonian University,\\ 
and Copernicus Center for Interdisciplinary Studies,\\ 
Orla 171,  Krak\'ow 30--244, Poland,\\
email: lech.sokolowski@uj.edu.pl,\\
email: zdzislaw.golda@uj.edu.pl} 

\date{}
\maketitle

\begin{abstract}
This is a consecutive paper on the timelike geodesic structure of static spherically symmetric 
spacetimes. First we show that for a stable circular orbit (if it exists) in any of these 
spacetimes all the infinitesimally close to it timelike geodesics constructed with the aid of 
the general geodesic deviation vector have the same length between a pair of conjugate points. 
In Reissner--Nordstr\"{o}m black hole metric we explicitly find the Jacobi fields on the radial 
geodesics and show that they are locally (and globally) maximal curves between any pair of their 
points outside the outer horizon. If a radial and circular geodesics in R--N metric have 
common endpoints, the radial one is longer. If a static spherically symmetric spacetime is 
ultrastatic, its gravitational field exerts no force on a free particle which may stay at rest; 
the free particle in motion has a constant velocity (in this sense the motion is uniform) 
and its total energy always exceeds the rest energy, i.~e.~it has no gravitational energy. 
Previously the absence of the gravitational force has been known only for the global 
Barriola--Vilenkin monopole. In the spacetime of the monopole we explicitly find all timelike 
geodesics, the Jacobi fields on them and the condition under which a generic geodesic may 
have conjugate points.\\



\end{abstract}

\section{Introduction}
This work is fourth in a series of papers~\cite{S, SG1, SG2} on the geodesic structure 
of various spacetimes. These investigations have originated from recent interest in the twin 
paradox in curved spacetimes \cite{Io1, Io2, JW, DG, ABK, BMW}. 
As is well known there is no paradox at all and there is a purely geometrical problem of which 
curve of all timelike paths connecting two given spacetime points is the longest one. Actually 
the problem consists of two different problems. Firstly, one is interested in the problem of 
determining the locally maximal curve. Assume that two given points, $p$ and $q$, are 
connected by a timelike geodesic $\gamma(0)$ and let us take a bundle of nearby timelike curves 
from $p$ to $q$, that is the curves which are everywhere at $\varepsilon$--distance from 
$\gamma(0)$. Then one seeks for the longest curve in the bundle. A general formalism for 
dealing with this local maximality problem was presented in the monograph \cite{HE} and the 
most relevant theorems were quoted in \cite{S}. The method of solving it is algorithmic: one 
solves the geodesic deviation equation on $\gamma(0)$ and finds the points at which the 
geodesic deviation vector field vanishes, these are conjugate points on the geodesic. If the 
segment of $\gamma(0)$ from $p$ to $q$ contains no points conjugate to $p$ the geodesic is 
locally the longest curve. In practice one expands the deviation vector field in a suitably 
chosen basis of three spacelike mutually orthogonal and parallelly transported along 
$\gamma(0)$ vector fields, then the geodesic deviation equation containing second absolute 
derivatives of the deviation vector is replaced by three equations for the Jacobi scalars (the 
coefficients of the expansion).\\
Yet the global problem concerns finding out the longest curve among all possible timelike curves 
joining $p$ and $q$ and one should take into account curves which are arbitrarily far from 
each other (besides the endpoints). Clearly the problem is different from the local one because 
it is not algorithmic: there is no effective method allowing one to determine in a finite 
number of steps the globally maximal worldline. In \cite{SG1} we briefly show what is known on 
the subject by quoting the most relevant `existence theorems' from the monograph \cite{BEE}. 
It turns out that a timelike geodesic is not globally maximal beyond the future cut point 
which in the problem takes place of the conjugate point. If the spacetime admits an isometry, 
such as spherical symmetry, it is possible to determine globally maximal timelike geodesics 
between pairs of points which are distinguished by the isometry.\\
Our investigations here are heavily based on the formalisms developed in \cite{SG1} and 
partially in \cite{S}. The paper is organised as follows. In section 2 we deal with one aspect 
of the problem of the locally maximal worldlines. A future directed timelike geodesic from 
$p$ to $q$ is locally maximal if the first future conjugate point to $p$ lies beyond the 
segment $pq$. Yet to the best of our knowledge there are no theorems stating what occurs if $q$ is conjugate to $p$ besides the fact that the geodesic $\gamma(0)$ is not the unique one 
locally maximal; the question is whether the geodesics $\gamma(\varepsilon)$ nearby to 
$\gamma(0)$ are of equal length or not. Here (sect.~2) we show that in the case of static spherically symmetric spacetimes and circular geodesics all the nearby geodesic curves determined by the general geodesic deviation vector field have the same length as the circular one from any $p$ to the first conjugate point $q$. One is unable to determine conjugate points (and possibly cut points) on a generic timelike geodesic in the given spacetime  due to technical difficulties: the geodesic deviation equation is intractable in the general case, even for high symmetries. Usually the investigation must be restricted to geometrically distinguished geodesic lines, such as radial or circular ones (if exist). Therefore in section 3 we study radial and circular geodesics in the Reissner--Nordstr\"{o}m black hole metric. An exception to the restriction is provided by a narrow class of Lorentzian manifolds, the ultrastatic spherically symmetric spacetimes which are so simple that allow one to explicitly find generic timelike geodesic curves and furthermore to explicitly solve (in the form of some 
integrals) the equations for the Jacobi scalars along these lines. In section 4 we develop 
a general formalism for these spacetimes. The formalism is applied in section 5 to the spacetime of the global Barriola--Vilenkin monopole without mass. Brief conclusions are contained in sect.~6.

For concreteness and as a trace of the original twin paradox we assume that a circular 
geodesic is followed by the twin B and the radial one is the worldline of the twin C. We 
consider only timelike geodesics and do not mark this fact each time. We use all the concepts and conventions as in \cite{SG1} and \cite{S}.

\section{Properties of timelike geodesics which intersect the timelike circular geodesics in two points in static spherically symmetric spacetimes}

In \cite{SG1} we investigated the conjugate points for timelike circular geodesic curves in a general 
static spherically symmetric (SSS) spacetime with the metric in the standard coordinates 
\begin{equation}\label{n1}
ds^2=e^{\nu(r)}\,dt^2-e^{\lambda(r)}\,dr^2-r^2(d\theta^2+\sin^2\theta\,d\phi^2),
\end{equation} 
for arbitrary functions $\nu(r)$ and $\lambda(r)$. It was found that a timelike circular geodesic 
exists at $r=r_0$ if and only if $\nu'_0\equiv d\nu/dr (r_0)>0$ and $r_0\,\nu'_0<2$. In other terms, if 
$g_{00}=e^{\nu}$ is a decreasing function (e.~g.~in de Sitter space), circular particle orbits do not 
exist. On physical grounds one is interested in stable particle orbits and the orbit $r=r_0$ is stable 
if the effective potential reaches minimum on it, what amounts to \cite{SG1}
\begin{equation}\label{n2}
\nu''_0-\nu_0'{}^{2}+\frac{3\nu'_0}{r_0}>0
\end{equation} 
and this means that $r_0$ is larger than the radius $r_I$ of the innermost stable circular orbit (ISCO). 
It is also clear that an unstable geodesic (circular or not) cannot contain conjugate points to any of 
its points and the explicit form of the deviation vectors confirms this expectation in the case of 
circular orbits.

Let $r=r_0>r_I$ be a stable circular orbit denoted by B. As it was shown in \cite{SG1} there are in general 
three infinite sequences of points conjugate to each point of the curve (assuming that the geodesic is 
infinitely extended to the future). Each sequence determines a continuous set of timelike geodesics 
intersecting B at points conjugate to an arbitrary initial point $P_0$. Here we shall compute the lengths 
of geodesics in each of the three sets from $P_0$ to a conjugate point. Let B be parameterized by its 
arc length $\tau$ (we shall denote by $s$ the arc length of the geodesics intersecting B at the conjugate 
points) and, as usual, the angular coordinates are so chosen that the spatial circle of B lies in the 2--surface $\theta=\pi/2$, then its equation is $x^{\alpha}=x_0^{\alpha}(\tau)$ with 
	\begin{equation}\label{n3}
t-t_0=\frac{k}{\kappa}e^{-\nu_0}\,\tau \qquad \textrm{and} \qquad \phi-\phi_0=\frac{L}{r_0^2}\,\tau,
	\end{equation}  
where
	\begin{equation}\label{n4}
\frac{k^2}{\kappa^2}=\frac{2e^{\nu_0}}{2-r_0\,\nu'_0} \qquad \textrm{and} \qquad L^2=\frac{r_0^3\nu'_0}
{2-r_0\nu'_0}.
	\end{equation} 
Here $k$ is the conserved (dimensionless) energy per unit particle's mass determined by the unique 
timelike Killing vector (and $\kappa$ is the normalization factor of the Killing field) and $L$ is the 
conserved angular momentum. An infinitesimal deviation vector $\varepsilon Z^{\alpha}(\tau)$ (Jacobi 
vector field) being any solution of the geodesic deviation equation on B connects the point 
$x_0^{\alpha}(\tau)$ on B to the corresponding point $x^{\alpha}(\tau)$ on an infinitesimally close 
geodesic $\gamma(\varepsilon)$. In other terms, the nearby geodesic $\gamma(\varepsilon)$ is parameterized by the arc length of B,
	\begin{equation}\label{n5}
x^{\alpha}(\tau)=x_0^{\alpha}(\tau)+\varepsilon Z^{\alpha}(\tau), \qquad |\varepsilon|\ll 1.
	\end{equation} 
Whereas the functions $x_0^{\alpha}(\tau)$ exactly represent a timelike geodesic, the coordinates 
$x^{\alpha}(\tau)$ satisfy the geodesic equation merely in the linear approximation in $\varepsilon$. 
In this sense the curve $\gamma(\varepsilon)$ is regarded as a geodesic one and if emanates from 
$P_0$ on B, then it intersects B at points conjugate to $P_0$. According to the well known theorem (cited in \cite{S}) a timelike geodesic has the locally maximal length between points $P_0$ and $P_1$ 
if and only if there is no point conjugate to $P_0$ in the open segment $P_0P_1$, otherwise there is a nearby timelike curve from $P_0$ to $P_1$ which is longer than the geodesic. Here we compare the lengths of the geodesic curves $\gamma(\varepsilon)$ with the length of B from $P_0$ to the first conjugate point.

We first generally show that the variation of the geodesic length is of order $\varepsilon^2$. Let $\gamma_0=\gamma(0)$ be any timelike geodesic parameterized by its arc length $\tau$, $x_0^{\alpha}=x_0^{\alpha}(\tau)$, and let $\gamma(\varepsilon)$ be any timelike (not necessarily geodesic) infinitesimally close curve also parameterized by $\tau$, given by eq.~(\ref{n5}), where $Z^{\alpha}(\tau)$ 
is a vector field (not necessarily a Jacobi field) orthogonal to $\gamma_0$, i.~e. 
$g_{\alpha\beta}u_0^{\alpha}\,Z^{\beta}=0$ along $\gamma_0$; here $u_0^{\alpha}=dx_0^{\alpha}/d\tau$ 
is the unit tangent vector to $\gamma_0$. The vector tangent to $\gamma(\varepsilon)$ is 
	\begin{equation}\label{n6}
u^{\alpha}=\frac{dx_0^{\alpha}}{d\tau}+\varepsilon\,\frac{dZ^{\alpha}}{d\tau}\equiv u_0^{\alpha}+
\varepsilon\dot{Z}^{\alpha}.
	\end{equation} 
Denoting by $D/d\tau$ the absolute derivative along $\gamma_0$ one finds
	\begin{eqnarray}\label{n7}
\frac{d}{d\tau}\left(g_{\alpha\beta}u_0^\alpha Z^\beta\right)&=&
\frac{D}{d\tau}\left(g_{\alpha\beta}u_0^\alpha Z^\beta\right)=
g_{\alpha\beta}u_0^\alpha \frac{D}{d\tau}Z^\beta =
g_{\alpha\beta}u_0^\alpha\left(\dot{Z^\beta}+\Gamma_{\mu\nu}^\beta u_0^\mu Z^\nu \right)\nonumber\\
&=&g_{\alpha\beta}u_0^\alpha \dot{Z^\beta}+\frac{1}{2}g_{\alpha\beta,\mu}u_0^\alpha u_0^\beta Z^\mu=0,
	\end{eqnarray} 
where we have used the symmetric Levi--Civita connection. On the other hand the squared length of the 
tangent vector $u^{\alpha}$ on $\gamma(\varepsilon)$ is 
	\begin{eqnarray}\label{n8}
g_{\alpha\beta}(x^{\mu}) u^\alpha u^\beta&=&g_{\alpha\beta}(x_0^\mu+\varepsilon Z^\mu)
(u_0^\alpha+\varepsilon \dot{Z^\alpha})(u_0^\beta+\varepsilon \dot{Z^\beta})\cong {}\nonumber\\
&\cong {}&1+\varepsilon (2 g_{\alpha\beta}(x_0)u_0^\alpha \dot{Z^\beta}+g_{\alpha\beta,\mu}Z^\mu u_0^\alpha u_0^\beta)\nonumber\\
&+&\varepsilon^2(g_{\alpha\beta}(x_0)\dot{Z^\alpha} \dot{Z^\beta}
+2 g_{\alpha\beta,\mu}u_0^\alpha \dot{Z^\beta} Z^\mu+\frac{1}{2}g_{\alpha\beta,\mu\nu}u_0^\alpha u_0^\beta Z^\mu Z^\nu)\nonumber\\
	\end{eqnarray} 
and applying (\ref{n7}) one sees that $u^{\alpha}u_{\alpha}$ differs from 1 by a term of order $\varepsilon^2$, $u^{\alpha}u_{\alpha}=1+\varepsilon^2 W$ with some scalar function $W$.

We now return to the circular geodesic B in an SSS spacetime, $r=r_0$. Its length from $P_0$ to the first conjugate point at distance $\tau_c$ is by definition $L(B)=\tau_c$, whereas the length of each geodesic $\gamma(\varepsilon)$ from $P_0$ to that point is 
	\begin{equation}\label{n9}
L(\gamma(\varepsilon))=\int^{\tau_c}_0 \left[g_{\mu\nu}(x^{\alpha})\frac{dx^{\mu}}{d\tau}
\frac{dx^{\nu}}{d\tau}\right]^{1/2}\,d\tau\cong \tau_c+\frac{1}{2}\varepsilon^2\int^{\tau_c}_0 
W\,d\tau.
	\end{equation} 
1. The simplest deviation field on B is $Z^{\alpha}=C\sin(L\tau/r_0^2)(0,0,1,0)$ with arbitrary 
dimensionless constant $C$ and $\gamma(\varepsilon)$ is given by 
	\begin{equation}\label{n10}
t-t_0=\frac{k}{\kappa}e^{-\nu_0}\tau, \quad r=r_0, \quad \theta=\frac{\pi}{2}+\varepsilon\,
C\sin\left(\frac{L\tau}{r_0^2}\right), \quad \phi-\phi_0=\frac{L}{r_0^2}\tau.
	\end{equation} 
The points conjugate to $P_0(t_0,r_0,\pi/2,\phi_0)$ are at equal distances $\tau_n=n\pi r_0^2/L$, $n=1,2,\ldots$ and from (\ref{n3}) one sees that they are located in space at $\phi_n=\phi_0+n\pi$. The first conjugate point $Q_1$ is antipodal to $P_0$ in the 2--surface $\theta=\pi/2$. Geometrically this is obvious, for if one rotates in the space the surface by any angle about the axis joining the points $\phi=\phi_0$ and $\phi=\phi_0+\pi$ one gets a circular geodesic intersecting B at these two points. By symmetry one expects that each geodesic (including the nearby $\gamma(\varepsilon)$) formed in 
this way has the same length as B. In fact, applying (\ref{n10}), (\ref{n8}) and (\ref{n9}) one gets $L(\gamma(\varepsilon))=L(B)=\tau_1=\pi r_0^2/L$. \\
2. Next we take a special solution to the geodesic deviation equation depending on one integration constant, which generates a family of close geodesics $\gamma(\varepsilon)$ whose spatial orbits also lie in the surface $\theta=\pi/2$. These are given by~\cite{SG1}
	\begin{eqnarray}\label{n11}
t-t_0 & = & \frac{k}{\kappa}e^{-\nu_0}\tau-\varepsilon TA(1-\cos y),\nonumber\\
r-r_0 & = & \frac{1}{2}\varepsilon XA\sqrt{4-b}\sin y, \qquad \theta=\frac{\pi}{2},\nonumber\\
\phi-\phi_0 & = & \frac{L}{r_0^2}\tau-\varepsilon YA(1-\cos y),
	\end{eqnarray} 
where $y\equiv \sqrt{4-b}\,q\tau$ and $T$, $X$, $Y$, $b$ and $q$ are constants determined by the values of $r_0$, $\nu_0'$, $\nu''_0$ and $\lambda_0$. For the stable circular orbits the condition (\ref{n2}) implies $b<4$. The curves of this family intersect B at the conjugate points $Q'_n(\tau'_n)$, $n=1,2,\ldots$, where 
	\begin{equation}\label{n12}
\tau'_n=\frac{2n\pi}{q\sqrt{4-b}}=2n\pi\left(\frac{r_0(2-r_0\nu'_0)e^{\lambda_0}}{3\nu'_0+
r_0\nu''_0-r_0\nu_0'{}^{2}}\right)^{1/2}.
	\end{equation}
For $\gamma(\varepsilon)$ the scalar $W$ defined in (\ref{n9}) is 
	\begin{equation}\label{n13}
W=-A^2\frac{e^{-\lambda_0}}{2r_0\nu'_0}(2-r_0\nu'_0)^{-2}(3\nu'_0+r_0\nu''_0-r_0\nu_0'{}^{2})^2 \cos 2y
\equiv -D^2\cos 2y
	\end{equation}
and its length from $P_0$ to $Q'_1$ is 
	\begin{equation}\label{n14}
L(\gamma(\varepsilon))=\tau'_1-\frac{1}{2}\varepsilon^2 D^2\int^{2\pi}_0 \cos 2y\,\frac{dy}
{q\sqrt{4-b}}=\tau'_1=L(B).
	\end{equation}
3. Finally we consider the most general Jacobi vector field depending on two integration constants 
$A_1$ and $A_4$ and again giving rise to orbits in the 2--surface $\theta=\pi/2$. Since the Jacobi 
field is determined up to a constant factor, we can put $A_1=2$ and the modified value of $A_4$ 
denote by $A$. (The solution (\ref{n11}) corresponds to $A_4=0$ and $A_1\equiv A\neq 0$.) Then $\gamma(\varepsilon)$ is \cite{SG1} 
	\begin{eqnarray}\label{n15}
t-t_0 & = & \frac{k}{\kappa}e^{-\nu_0}\tau+\varepsilon TF(y),
\nonumber\\
r-r_0 & = & \varepsilon X\sqrt{4-b}[\sin y+A(\cos y-1)], \qquad \theta=\frac{\pi}{2},
\nonumber\\
\phi-\phi_0 & = & \frac{L}{r_0^2}\tau+\varepsilon YF(y)
	\end{eqnarray}
with
\begin{equation}\label{n16}
F(y)\equiv 2(\cos y-1)+A\left(\frac{1}{2}by-2\sin y\right).
\end{equation}
As previously $y=\sqrt{4-b}q\tau$. If $\gamma(\varepsilon)$ is to intersect B at some $y>0$, then two equations should hold, $A(\cos y-1)+\sin y=0$ and $F(y)=0$. These have an infinite sequence of roots, $y_n(b)$, $n=1, 2,\ldots$, and each root is a solution of 
	\begin{equation}\label{n17}
\cos y +\frac{b}{8}y\sin y -1=0
	\end{equation}
and in consequence it determines $A$,
	\begin{equation}\label{n18}
A(n,b)=\frac{\sin y_n(b)}{1-\cos y_n(b)}.
	\end{equation}
This means that one actually has a two--parameter family of geodesics $\gamma(\varepsilon,n,b(r_0))$ close to the circular B. Each geodesic of this family emanates from $P_0(\tau=0)$ on B, 
spatially belongs to the 2--surface $\theta=\pi/2$ and intersects B only once at $\bar{Q}_n
(\bar{\tau}_n)$, 
	\begin{equation}\label{n19}
\bar{\tau}_n=\frac{y_n(b(r_0))}{q\sqrt{4-b(r_0)}}=\left(\frac{r_0(2-r_0\nu'_0)e^{\lambda_0}}{3\nu'_0+r_0\nu''_0-r_0\nu_0'{}^{2}}\right)^{1/2}\,y_n(b).
	\end{equation}
In fact, for given $A(n,b)$ the two equations have no roots other than $y_n(b)$. By expanding 
eq.~(\ref{n17}) into a power series around $y=(2n+1)\pi$ one finds an analytic approximate expression for $y_n(b)$. The first few roots are well approximated by 
\begin{equation}\label{n20}
y_n(b)=\frac{(2n+1)^2\pi^2 b-16}{(2n+1)\pi b}, \qquad n<10,
\end{equation}
whereas for large $n$ a good approximation is 
	\begin{equation}\label{n21}
y_n=(2n+1)\pi-\delta_n(b), \qquad \delta_n(b)=\frac{16(2n+1)\pi}{(2n+1)^2\pi^2 b-16}.
	\end{equation}
For example, for $b=3.6$ the difference between the numerically found value $y_{15}=97.3437$ and the analytic value given by (\ref{n21}) is of order $10^{-5}$. The geometrical interpretation of the origin of these conjugate points is unclear.\\ 
One computes the length of $\gamma(\varepsilon,n,b(r_0))$ between $P_0$ and $\bar{Q}_n(\bar{\tau}_n)$ applying (\ref{n15}), (\ref{n16}) and (\ref{n9}),
	\begin{equation}\label{n22}
L(\gamma(\varepsilon,n,b))\cong \bar{\tau}_n+\varepsilon^2\frac{\nu_0'}{16r_0 q} (4-b)^{3/2}
e^{-\lambda_0}\left(\!\frac{\sin y_n}{1-\cos y_n}\!\right)^2
\left(\! b y_n-\frac{8(1-\cos y_n)}{\sin y_n} \!\right)
	\end{equation}
By applying eq.~(\ref{n17}) one immediately sees that the last round bracket 
vanishes implying that for each geodesic of this family its length between the two conjugate 
points $P_0$ and $\bar{Q}_n$ is also equal to the length of the circular geodesic B, 
$L(\gamma(\varepsilon,n,b))=L(B)=\bar{\tau}_n$.

One concludes from these three cases that at least in the case of circular geodesics in SSS 
spacetimes all the nearby geodesic curves determined by the infinitesimal deviation vector 
fields have the same length from any initial point to the first future conjugate point to it, 
equal to the length of the circular geodesic. It is well known (see Theorem 1 cited in~\cite{SG1}) that the future cut point of $P_0$ comes no later than the first future 
conjugate point to $P_0$. Whether or not there exist distant timelike geodesics joining the 
two points which are longer than the circular B remains an open problem. A definite answer 
is known in the case of the first sequence of conjugate points on B if one asks about 
geodesic lengths to the second conjugate point. In this case the second conjugate point $Q_2$ coincides in the space with $P_0$ (B makes one revolution) and in Schwarzschild metric it is known \cite{S} that the radial geodesic directed outwards and returning to $r=r_0$ 
simultaneously with B is longer than the latter. We will see in the next section that the 
same holds in Reissner--Nordstr\"{o}m spacetime.

\section{Reissner--Nordstr\"{o}m spacetime}

We assume that the spacetime represents a nonrotating charged black hole, then the metric is 
	\begin{equation}\label{n23}
ds^2=\left(1-\frac{2M}{r}+\frac{Q^2}{r^2}\right)dt^2-\left(1-\frac{2M}{r}+\frac{Q^2}{r^2}\right)^{-1}dr^2-r^2d\Omega^2
	\end{equation}
($c=G=1$) with the mass parameter exceeding the charge, $M^2>Q^2$. The timelike Killing vector $K^{\alpha}=\delta_0^{\alpha}$, normalized to $+1$ at the spatial infinity, actually remains timelike in the region where $g_{00}=e^{\nu}>0$, i.~e.~for $r>r_+$ and for $r<r_-$, where $r_{\pm}=M\pm\sqrt{M^2-Q^2}$. We shall deal only with the region outside the outer event horizon, $r>r_+=M+\sqrt{M^2-Q^2}$. The equations for a timelike geodesic reduce to an equation for the radial coordinate,
	\begin{equation}\label{n24}
-\ddot{r}+\frac{Mr-Q^2}{r(r^2-2Mr+Q^2)}(\dot{r}^2-k^2)+\frac{L^2}{r^3}\left(1-\frac{2M}{r}+
\frac{Q^2}{r^2}\right)=0
	\end{equation}
($\dot{f}\equiv df/ds$) plus the integral of energy $k$ per unit particle's mass, $k\equiv 
E/(mc^2)$, $E\equiv mc^2 \dot{x}^{\alpha}K_{\alpha}$, giving rise to 
	\begin{equation}\label{n25}
\dot{t}=k\left(1-\frac{2M}{r}+\frac{Q^2}{r^2}\right)^{-1}
	\end{equation}
and the conserved angular momentum $L$, $\dot{\phi}=L/r^2$. The motion is `flat', $\theta=\pi/2$, 
and the universal integral of motion, $g_{\alpha\beta}\dot{x}^{\alpha}\dot{x}^{\beta}=1$, 
allows one to replace eq. (\ref{n24}) by the following first order equation 
\begin{equation}\label{n26}
\dot{r}^2=k^2-\left(\frac{L^2}{r^2}+1\right)\left(1-\frac{2M}{r}+\frac{Q^2}{r^2}\right).
\end{equation}
We first consider radial geodesics.

\subsection{Radial timelike geodesics}

We assume that the radial timelike geodesic representing the worldline of the twin C emanates 
from the event $P_0(t=t_0,r=r_0>r_+,\theta=\pi/2, \phi=\phi_0)$ outwards with initial velocity 
$\dot{r}(t_0)=u>0$, reaches the maximal height $r=r_M$ at $t=t_M$, turns down and at $P_1(t=t_1)$ returns to the initial point $r=r_0$ and then goes downwards to the outer horizon $r=r_+$. We do not follow farther the geodesic which actually crosses the horizon since it would require the appropriate change of the chart. The integral of motion (\ref{n26}) is reduced to 
	\begin{equation}\label{n27}
\dot{r}^2=k^2-1+\frac{2M}{r}-\frac{Q^2}{r^2}
	\end{equation}
and one sees that $\dot{r}\geq 0$ for $r\to \infty$ if $k\geq 1$, what means that (as in 
Schwarzschild spacetime) if the total energy (kinetic, potential and rest mass one) exceeds the rest energy, the twin C may escape to the spatial infinity and will not return. We therefore 
assume that the energy of the geodesic C is $0<k<1$. Then the height of the flight is 
	\begin{equation}\label{n28}
r_M=\frac{1}{1-k^2}\left(M+\sqrt{M^2-(1-k^2)Q^2}\right),
	\end{equation}
this is a monotonically growing function of $k$ from $r_M=r_0$ for $k^2=g_{00}(r_0)$ to infinity for $k\to 1$. Conversely, the energy may be expressed in terms of $r_M$ as $k^2=g_{00}(r_M)$. It is convenient to parameterize C on the entire arc from $P_0$ to the horizon by an angle $\eta$,
	\begin{equation}\label{n29}
r(\eta)=r_M\cos^2\eta\equiv \frac{1}{2}r_M(\cos 2\eta+1).
	\end{equation}
Then $\eta=-\alpha/2$ at $P_0$ and $\eta=+\alpha/2$ at $P_1$, or $r(\pm\alpha/2)=\frac{1}{2}
r_M(\cos\alpha+1)=r_0$ and it follows that $\cos\alpha=2r_0/r_M-1$ or $\cos^2\frac{\alpha}{2}=
r_0/r_M$. If $r_M=r_0$ then $\alpha=0$ and if $r_M\to\infty$ then $\alpha\to\pi$, hence 
$0\leq \alpha<\pi$. The initial point is $-\alpha/2<0$ since $r(0)=r_M$. On the ingoing segment 
of the geodesic $\eta>0$ and it reaches the outer horizon for $\eta_H$, $r(\eta_H)=r_+=
r_M\cos^2\eta_H$, what implies $\eta_H=\arccos(r_+/r_M)^{1/2}$ and since 
$\cos\frac{\alpha}{2}=\sqrt{\frac{r_0}{r_M}}>\sqrt{\frac{r_+}{r_M}}$ 
one finally gets $\eta\in[-\frac{\alpha}{2},\eta_H)$ where $0\leq\alpha/2<\eta_H<\pi/2$. The 
radial component of the velocity $\dot{r}$ may be expressed from (\ref{n27}) in terms of $\eta$ as 
	\begin{equation}\label{n30}
\frac{dr}{ds}=-\left(1-k^2-\frac{Q^2}{r_M^2\cos^2\eta}\right)^{1/2}\,\tan\eta.
	\end{equation}
The length of the geodesic C from $P_0$ to any point $\eta$ may be computed as follows. The metric along this curve is $ds^2=e^{\nu}\,dt^2-e^{-\nu}\,dr^2$ and one finds from it that 
	\begin{equation}\label{n31}
\left(\frac{ds}{d\eta}\right)^2=e^{-\nu}\left(\frac{dr}{d\eta}\right)^2\left[e^{\nu}\left(\frac{dt}
{ds}\right)^2-1\right]^{-1}.
	\end{equation}
Then applying (\ref{n25}) and (\ref{n29}) one gets after some manipulations that for both the outgoing and ingoing segments of C
	\begin{equation}\label{n32}
\frac{ds}{d\eta}=2\left(\frac{r_M^3}{M}\right)^{1/2}[(2-\beta)\cos^2\eta-\beta]^{-1/2}\,
\cos^3\eta,
	\end{equation}
where $\beta\equiv Q^2/(Mr_M)<1/2$. This expression is integrated out and the outcome is expressed in terms of the function 
	\begin{equation}\label{n33}
F(\beta,\eta)\equiv (2-\beta)\cos^2\eta-\beta 
	\end{equation}
as
	\begin{eqnarray}\label{n34}
&&s(\eta)=
\left(\frac{r_M^3}{M}\right)^{1/2}(2-\beta)^{-3/2}
\left[
2\arctan\left[
\sqrt{\frac{2-\beta}{F(\beta,\eta)}}\sin\eta
\right]+\right.\nonumber\\
&&\quad{}+2\arctan \left[ \sqrt{\frac{2-\beta}{F(\beta,\frac{\alpha}{2})}}\sin\frac{\alpha}{2}\right]\nonumber\\
&&+\sqrt{2-\beta}\left.\left[\sqrt{F(\beta,\eta)}\sin\eta +\sqrt{F\left(\beta,\frac{\alpha}{2}\right)}\sin\frac{\alpha}{2}\,
\right]\right]
	\end{eqnarray}
and this formula immediately gives the length of C from $P_0$ to $P_1$,
	\begin{eqnarray}\label{n35}
s_C=s\left(\frac{\alpha}{2}\right)&=&\left(\frac{r_M^3}{M}\right)^{1/2}(2-\beta)^{-3/2}
\left[4\arctan\left[
\sqrt{\frac{2-\beta}{F\left(\beta,\frac{\alpha}{2}\right)}}
\sin\frac{\alpha}{2}
\right]\right.+{}\nonumber\\
&&{}+2\sqrt{2-\beta}\left.\sqrt{F\left(\beta,\frac{\alpha}{2}\right)}\sin\frac{\alpha}{2}\,
\right].
	\end{eqnarray}
The last two formulae make sense if $F>0$ in the whole interval $-\alpha/2\leq\eta<\eta_H$. This requires $\cos^2\eta>\beta/(2-\beta)$ in the interval and since the lowest value of $\cos^2\eta$ is for $\eta=\eta_H$ one gets $r_+/r_M>\beta/(2-\beta)$. The ratio of these two quantities is 
	\begin{displaymath}
\frac{r_+}{r_M}\left(\frac{\beta}{2-\beta}\right)^{-1}=r_+\left(\frac{2M}{Q^2}-
\frac{1}{r_M}\right)
	\end{displaymath}
and applying $r_+>M$, $2M/Q^2>2/M$ and $1/r_M<1/M$ one finds that the ratio is greater than 1 showing that $F>0$ along the geodesic up to the horizon.

Our aim is to compare the length of C between $P_0$ and $P_1$ with the length of the circular geodesic connecting these two points in the spacetime. The coordinate time of making one circle $\Delta t_B$ on the circular geodesic is uniquely determined by its radius $r_0$, yet the time of flight on the radial C from $r_0$ to $r_M$ and back to $r_0$ depends on $r_M$ (or equivalently on the energy $k$). Thus if the twin B following the circular geodesic and the twin C on the radial curve start from $P_0$ and are to meet again at $P_1$ the height of flight $r_M$ must be precisely tuned to make the flight duration equal to $\Delta t_B$. To this end one expresses the time coordinate on C as a function of $\eta$. From (\ref{n25}) and (\ref{n32}) one gets 
	\begin{equation}\label{n36}
\frac{dt}{d\eta}=2k\left(\frac{r_M^5}{M^3}\right)^{1/2}\left(\frac{r_M}{M}\cos^4\eta-2\cos^2\eta+
\beta\right)^{-1}[F(\beta,\eta)]^{-1/2}\cos^7\eta.
	\end{equation}
It is quite surprising and fortunate that $t(\eta)$ is expressed in terms of elementary functions. 
By substitution $x\equiv \cos^2\eta$ the indefinite integral of eq.~(\ref{n36}) is reduced to 
\begin{equation}\label{n37}
-\varepsilon k\left(\frac{r_M^5}{M^3}\right)^{1/2}\int\frac{x^3\,[F(\beta,x)]^{-1/2}}
{\frac{r_M}{M}x^2-2x+\beta}\,\frac{dx}{\sqrt{1-x}}\equiv \varepsilon J(x),
\end{equation}
where $\varepsilon=-1$ for $\sin\eta<0$ and $\varepsilon=+1$ for $\sin\eta>0$. Calculating this 
integral requires considerable amount of ingenious work supported by the program Mathematica and the 
outcome reads 
	\begin{eqnarray}\label{n38}
J(x) & = & -k\left(\frac{r_M^5}{M^3}\right)^{1/2}\left\{-\frac{g(x)}{2\frac{r_M}{M}-\mu}+
\frac{1}{\sigma}(2-\beta)^{-3/2}\arcsin\left[\frac{(2-\beta)x-1}{1-\beta}\right]-\right.
\nonumber\\
& - & \frac{2M^2}{\sqrt{2-\beta}}\frac{(\sigma+\beta-2)}{r_M^2\sigma}\arctan\left[
\frac{(2-\beta)(1-x)}{(2-\beta)x-\beta}\right]^{1/2}+
\nonumber\\
&& \left.{}+a_+b_+\ln\left|\frac{f(x)+a_+}{f(x)-a_+}\right|-a_-b_-\ln\left|
\frac{f(x)+a_-}{f(x)-a_-}\right|\right\}.
	\end{eqnarray}
Here 
	\begin{eqnarray}\label{n39}
g(x) & = & \sqrt{1-x}\,\sqrt{(2-\beta)x-\beta}=\sqrt{(1-x)F(\beta,x)},
\nonumber\\
f(x) & = & \left(\frac{1-x}{F(\beta,x)}\right)^{1/2},\qquad \mu=\frac{Q^2}{M^2}, \quad 
\sigma=\frac{r_M}{M}+\beta-2,
\nonumber\\
a_{\pm} & = & \frac{1}{\beta\sqrt{\sigma}}[2(1-\beta)(1\pm\sqrt{1-\mu})-\mu+\beta^2]^{1/2}, 
\nonumber\\
b_{\pm} & = & \frac{\beta^3(1-\beta\pm\sqrt{1-\mu})}{2\sqrt{1-\mu}[(2-\beta)(1\pm\sqrt{1-\mu})-
\mu]^2},
	\end{eqnarray}
all the functions are well defined for $\beta/(2-\beta)<x\leq 1$ and this condition is 
satisfied along the geodesic from $\eta=-\alpha/2$ up to $r_M$ and down to the outer horizon. The time lapse to any $\eta>0$ is then 
$t(\eta)-t_0=[t(0)-t_0]+[t(\eta)-t(0)]=(-1)[J(1)-J(\cos^2\alpha/2)]+(+1)[J(\cos^2\eta)-J(1)]$. 
We need the time of flight to $r_M$ and back to $r_0$,
	\begin{equation}\label{n40}
t\left(\frac{\alpha}{2}\right)-t_0=2J\left(\cos^2\frac{\alpha}{2}\right)-2J(1),
	\end{equation}
where
	\begin{equation}\label{n41}
J(1)=-\frac{\pi}{2}k\frac{1}{\sigma}\left(\frac{r_M^5}{M^3}\right)^{1/2}(2-\beta)^{-3/2}.
	\end{equation}
We shall use this result in subsect.~3.3.

\subsection{The deviation vector fields on the radial timelike geodesics}

As mentioned in the Introduction, we replace the geodesic deviation equation for the deviation 
vector by three scalar equations for the Jacobi scalars $Z_a(s)$, $a=1,2,3$. The spacelike 
orthonormal basis triad orthogonal to the radial geodesic C is chosen as in \cite{SG1},
	\begin{equation}\label{n42}
e_1^{\mu}=[\varepsilon e^{-\nu}(k^2-e^{\nu})^{1/2}, k, 0,0], \quad e_2^{\mu}=\frac{1}{r}\,\delta
^{\mu}_2, \quad e_3^{\mu}=\frac{1}{r}\,\delta^{\mu}_3 
	\end{equation}
where $\varepsilon=+1$ on the outward directed segment of C and $\varepsilon=-1$ for the 
ingoing one; $r=r(\eta)$ according to (\ref{n29}). In terms of this basis the equations read 
	\begin{eqnarray}\label{n43}
\frac{d^2Z_1}{ds^2} & = & \frac{2Mr-3Q^2}{r^4}\,Z_1,
\nonumber\\
\frac{d^2Z_2}{ds^2} & = & \frac{Q^2-Mr}{r^4}\,Z_2, \qquad
\frac{d^2Z_3}{ds^2} =\frac{Q^2-Mr}{r^4}\,Z_3.
	\end{eqnarray}
Simplicity of these decoupled equations is deceptive, for their left hand sides are 
derivatives with respect to the proper time instead of $r$ or $\eta$. Replacing $d/ds$ by 
derivatives with respect to $\eta$ one arrives at 
	\begin{eqnarray}\label{n44}
F(\beta,\eta)\,\frac{d^2Z_1}{d\eta^2}+\frac{\sin 2\eta}{\cos 2\eta+1}[2F(\beta,\eta)-
\beta]\,\frac{dZ_1}{d\eta} & - &
\nonumber\\
\frac{8}{\cos 2\eta+1}(\cos 2\eta+1-3\beta)Z_1 & = & 0,
\nonumber\\
F(\beta,\eta)\,\frac{d^2Z_2}{d\eta^2}+\frac{\sin 2\eta}{\cos 2\eta+1}(2F(\beta,\eta)-
\beta)\,\frac{dZ_2}{d\eta} & - &
\nonumber\\
\frac{8}{\cos 2\eta+1}(\beta-\frac{1}{2}(\cos 2\eta+1))Z_2 & = & 0
	\end{eqnarray}
($F$ as in (33)) and the equation for $Z_3$ is identical with that for $Z_2$. The first 
integrals for these equations \cite{SG1} are generated by the timelike Killing vector 
$K_t^{\alpha}=\delta^{\alpha}_0$ and the three spacelike rotational Killing fields, which 
at the points of C are equal to 
	\begin{equation}\label{n45}
K_x^{\alpha}=(0,0,-\sin\phi_0,0), \qquad K_y^{\alpha}=(0,0,\cos\phi_0,0), \qquad 
K_z^{\alpha}=\delta^{\alpha}_3;
	\end{equation}
clearly $K^{\alpha}_x$ and $K_y^{\alpha}$ give rise to the same conserved quantity. Applying 
the general formalism \cite{SG1} one finds that $K_t^{\alpha}$ generates a first integral 
for $Z_1$, 
	\begin{equation}\label{n46}
F(\beta, \eta)\sin2\eta\,\frac{dZ_1}{d\eta}-2(\cos 2\eta+1-2\beta)Z_1=C_1(\cos 2\eta+1)^3,
\end{equation}
whereas $K^{\alpha}_x$ and $K_y^{\alpha}$ generate the same first integral for $Z_2$,
	\begin{equation}\label{n47}
\frac{dZ_2}{d\eta}+2Z_2\tan\eta=\frac{C_2\cos\eta}{[F(\beta,\eta)]^{1/2}},
\end{equation}
ultimately $K_z^{\alpha}$ generates the first integral for $Z_3$ which is identical with 
(47) for a different constant $C_3$. Clearly $C_1$, $C_2$ and $C_3$ are arbitrary constants. 
Equations (44) are linear homogeneous, hence their general solutions are\\
$Z_a(\eta)=C_{a1}\,Z_{a1}(\beta, \eta)+ C_{a2}\,Z_{a2}(\beta, \eta)$, $a=1,2,3$.\\
The equations (\ref{n44}) are symmetric (invariant) under the inversion $\eta\to -\eta$, also the 
first integral (\ref{n46})  is symmetric, whereas the first integrals for $Z_2$ and $Z_3$ are 
antisymmetric. This means that the special solutions $Z_{a1}$and $Z_{a2}$ have definite 
symmetry and if their symmetry agrees with that of the first integral (\ref{n46}) (resp. (\ref{n47})), then the solutions satisfy the first integral equations for some $C_a\neq 0$, otherwise they satisfy the latter for $C_a=0$.

1. Solutions for $Z_1$. The first solution is 
	\begin{equation}\label{n48}
Z_{11}(\beta, \eta)=\sqrt{F(\beta,\eta)}\,\frac{\sin\eta}{\cos^2\eta}=-Z_{11}(\beta,-\eta)
	\end{equation}
and (\ref{n46}) holds for $C_1=0$. The second solution is quite complicated,
	\begin{eqnarray}\label{n49}
Z_{12}(\beta, \eta)&=&\left\{6\sqrt{\frac{F(\beta,\eta)}{2-\beta}}\sin\eta\, 
\arctan\left[\sqrt{\frac{2-\beta}{F(\beta,\eta)}}\sin\eta\right]+{}\right.\nonumber\\
&&{}+ \frac{1}{(1-\beta)^2}\left[
-3\beta+4\beta^2-2\beta^3+(6-10\beta+5\beta^2)\cos^2\eta+\right. \nonumber\\ 
&&\left.\left.{}+(-2+5\beta-4\beta^2+\beta^3)\cos^4\eta
\right]\right\}\frac{1}{\cos^2\eta}
	\end{eqnarray}
and is symmetric, $Z_{12}(\beta,-\eta)=+Z_{12}(\beta,\eta)$ and it satisfies (\ref{n46}) for 
$C_1=-1$. In the limit $\beta\to 0$ ($Q^2\to 0$) one recovers the solutions valid for 
Schwarzschild metric \cite{S}.

2. The first solution for $Z_2$,
	\begin{equation}\label{n50}
Z_{21}=\cos^2\eta,
	\end{equation}
is independent of $\beta$, symmetric and satisfies (\ref{n47}) for $C_2=0$, clearly it also 
holds for Schwarzschild metric. Yet the second solution,
	\begin{equation}\label{n51}
Z_{22}(\beta, \eta)=\cos^2\eta\,\arctan\left(\sqrt{\frac{\beta}{F(\beta,\eta)}}\,\sin\eta
\right),
	\end{equation}
is antisymmetric, yields $C_2=\sqrt{\beta}$ in (\ref{n47}), is non--analytic for $\beta=0$ and 
in this limit tends to 0. (The corresponding special solution for $Q^2=0$ is just 
$\sin 2\eta$ \cite{S}.)

We now seek for points conjugate on the geodesic C to $P_0(\eta=-\alpha/2)$. Since the 
generic deviation vector is $Z^{\mu}=\sum_{a=1}^3\, Z_ae_a^{\mu}$, all the Jacobi scalars 
$Z_a$ should vanish both at $P_0$ and at the sought for conjugate points. In the search one 
sets two scalars identically zero and seeks for zeros of the third scalar.\\
a) First we set $Z_2=Z_3=0$. From (\ref{n48}) and (\ref{n49}) one finds that $Z_{11}(\beta,-\alpha/2)
\neq 0$ and $Z_{12}(\beta,-\alpha/2)\neq 0$. We set $C_{12}=1$, then the condition 
$Z_1(\beta,-\alpha/2)=C_{11}Z_{11}(\beta,-\alpha/2)+Z_{12}(\beta,-\alpha/2)=0$ 
determines $C_{11}(\beta,-\alpha/2)$ and we seek for roots of the function 
$Z_1(\beta, C_{11},\eta)$ for $\eta>-\alpha/2$. The numerical analysis shows that this 
$Z_1$ has no roots.\\
b) Let $Z_1=Z_3=0$. Since $Z_{21}$ and $Z_{22}$ are different from 0 for $\eta=-\alpha/2$ 
we set $C_{22}=1$ and analogously from the condition $Z_2(\beta,-\alpha/2)=0$ 
determine $C_{21}(\beta,-\alpha/2)$. Again the function $Z_2(\beta, C_{21},\eta)$ 
nowhere vanishes.\\
The case $Z_1=Z_2=0$ is identical to the second one. In this way we have shown that 
the radial timelike geodesic C has no conjugate points to the initial point $P_0$ 
for any $r_0>r_+$; this means that C is the locally maximal curve between any pair of 
its points. In \cite{SG1} we have shown that in R--N spacetime one can introduce the 
Gaussian normal geodesic (GNG) coordinates, i.~e.~the comoving ones, and the domain of 
these at least covers the same domain of the manifold as the standard coordinates 
$(t,r,\theta,\phi)$ do, i.~e.~from the outer horizon to the spatial infinity. In the 
GNG coordinates the radial geodesic C becomes the time coordinate line (the spatial 
coordinates are constant) and it is straightforward to prove \cite{SG1} that there is 
no timelike curve joining any pair of points on C whose length would be equal or 
greater than the length of the segment of C between these points. This means that 
each radial timelike geodesic is the globally maximal curve between its points.

\subsection{Circular timelike geodesics}

In this spacetime the general conditions od sect. 2 for the existence of circular 
geodesics reduce to $2-r_0\nu'_0>0$ since $\nu'$ is positive outside the outer 
horizon, $r>r_+$ (and for $M^2>Q^2$). The former condition implies 
$N\equiv r_0^2-3Mr_0+2Q^2>0$ and this amounts to 
	\begin{equation}\label{n52}
r_0>r_m\equiv \frac{3}{2}M\left(1+\sqrt{1-\frac{8Q^2}{9M^2}}\right),
	\end{equation}
where $2M<r_m<3M$. The value of $r_0$ determines all the parameters of the circular 
orbit, the energy $k$ and the angular momentum $L$. From (3)
	\begin{equation}\label{n53}
k^2=\frac{(r_0^2-2Mr_0+Q^2)^2}{r_0^2N}, \qquad L^2=\frac{r_0^2}{N}(Mr_0-Q^2)
	\end{equation}
and according to (3) the time and azimuthal coordinate are parameterized by the 
arc length of the geodesic B,
	\begin{equation}\label{n54}
t-t_0=ke^{-\nu_0}\,s=\frac{r_0}{\sqrt{N}}s, \qquad \phi-\phi_0=\frac{1}{r_0}
\left(\frac{Mr_0-Q^2}{N}\right)^{1/2}\,s.
	\end{equation}
The length of B corresponding to one revolution around the black hole is then 
	\begin{equation}\label{n55}
s_B=\frac{2\pi}{L}\,r_0^2=2\pi r_0\left(\frac{N}{Mr_0-Q^2}\right)^{1/2}
\end{equation}
and the corresponding lapse of the coordinate time is 
	\begin{equation}\label{n56}
\Delta t_B=t(s_B)-t_0=\frac{2\pi r_0^2}{(Mr_0-Q^2)^{1/2}}.
\end{equation}
If $r_0$ is very close to $r_m$ the geodesic is unstable. In R--N spacetime the 
radius $r_I$ of the ISCO is given by the unique real root of the cubic equation 
	\begin{equation}\label{n57}
J(r_I)\equiv r_I^3-6Mr_I^2+9Q^2r_I-4\frac{Q^4}{M}=0.
	\end{equation}
For fixed mass $M$ the values of $r_I(Q)$ are given in Table 1 as a function of the 
ratio $\mu\equiv Q^2/M^2$, $0<\mu<1$.
	\begin{table}[h]
\caption{The values of $r_I(Q)/M$ as a function of $\mu= Q^2/M^2$}
\label{Tabela01}
$$
\begin{tabular}{|c|c|}
	  \hline 
 $ \mu$ & $r_I/M$\\
	\hline
	\hline
  0 & 6 \\
	\hline
0.1&5.84725\\
\hline
0.2&5,6852\\
\hline
0.3&5.52293\\
\hline
0.4&5.34939\\
\hline
0.5&5.16646\\
\hline
0.6&4.97221\\
\hline
0.7&4.76392\\
\hline
0.8&4.53759\\
\hline
0.9&4.28678\\
\hline
1&4\\
\hline
\end{tabular} 
$$
	\end{table}
One sees that $r_I(Q)$ decreases approximately linearly from $6M$ to $4M$. We shall 
assume that the geodesic B is stable, $r_0>r_I(Q)$, then, as it was mentioned in sect. 2, 
B has infinite number of conjugate points ordered in three sequences.

We are now able to compare the lengths of the radial and the circular geodesic having 
common endpoints $P_0$ and $P_1$. The time interval between these two events is equal to 
$t(\alpha/2)-t_0$ if counted along the radial C and $\Delta t_B$ as necessary for one 
revolution arond the black hole. Equating the two intervals, (\ref{n40}) and (\ref{n56}),
	\begin{equation}\label{n58}
\frac{2\pi r_0^2}{(Mr_0-Q^2)^{1/2}}=2J(\cos^2\frac{\alpha}{2})-2J(1),
	\end{equation}
one gets an algebraic transcendental equation for the height $r_M$ of the radial flight. 
The roots of the equation depend on two parameters: $r_0$ and the ratio $\mu=Q^2/M^2$. 
The equation was numerically solved for three values of $\mu$; for each value of $\mu$ 
two values of $r_0$ were taken: one very close to the corresponding radius of the ISCO 
and the other was larger. The solutions $r_M(r_0,\mu)$ are unique (in the space of 
variables $r_0$, $r_M$ and $\mu$ eq.~(58) represents a surface which is almost exactly 
a plane) and are inserted into eq.~(35) for the length $s_C$ of the radial geodesic 
and this length is compared with the length $s_B$ of the circular one in (\ref{n55}). Table 2 
shows values of $r_M/r_0$ and $s_C/s_B$ for the chosen values of $r_0$ and $\mu$. 
	\begin{table}[h]
\caption{Six values of $r_M/r_0$ and $s_C/s_B$ for chosen values of $r_0/M$ and $\mu=Q^2/M^2$.}
\label{Tabela02}
$$
\begin{tabular}{|c|c|c|}
	  \hline
\multicolumn{3}{|c|}{$\mu=0.1$}\\
\hline
$r_0/M$&$r_M/r_0$&$s_C/s_B$\\
\hline
5.86		&2.07086&		1.23795\\
\hline
    10	&		2.14253&		1.10736\\
\hline
\multicolumn{3}{|c|}{$\mu=0.5$}\\
\hline
$r_0/M$&$r_M/r_0$&$s_C/s_B$\\
\hline
    5.17&		2.07478&		1.26553\\
\hline
    6	&		2.09954&		1.20797\\
\hline
  \multicolumn{3}{|c|}{$\mu=0.9$}\\
\hline
$r_0/M$&$r_M/r_0$&$s_C/s_B$\\
\hline
    4.3	&		2.07995&		1.4321\\
\hline
    10& 			2.17053&		1.11617\\
\hline
\end{tabular} 
$$
	\end{table}
One sees that $s_C$ is always larger than $s_B$ and their ratio slowly diminishes with 
growing $r_0$. A similar behaviour was found for the Schwarzschild black hole~\cite{S}.

\subsection{Jacobi fields and conjugate points on circular timelike geodesics}

For all SSS spacetimes all quantities and equations concerning stable circular 
geodesics are the same, only the constants appearing in the formalism depend in a 
different way on the radius $r_0$; we therefore refer the reader to \cite{SG1} for 
all definitions and equations. We only recall the orthonormal spacelike basis triad which 
is orthogonal to the circular geodesic B and parallelly transported along it,
	\begin{eqnarray}\label{n59}
e_1^{\mu} & = & [-T\sin qs, X\cos qs, 0, -Y \sin qs], \qquad e_2^{\mu}=[0,0,\frac{1}{r_0}, 0],
\nonumber\\
e_3^{\mu} & = & [T\cos qs, X\sin qs, 0, Y \cos qs].
	\end{eqnarray} 
The constants already appeared in (11) and now are equal to 
	\begin{equation}\label{n60}
T=\left[\frac{1}{N}e^{-\nu_0}(Mr_0-Q^2)\right]^{1/2},~X=e^{\nu_0/2},~~Y= \left[\frac{e^{\nu_0}}{N}\right]^{1/2},~~q^2=\frac{1}{r_0^4}(Mr_0-Q^2),
	\end{equation} 
furthermore in the equations for the Jacobi scalars $Z_a$ \cite{SG1} there appears the constant 
	\begin{equation}\label{n61}
b\equiv 3+\frac{1}{N}(3Mr_0-4Q^2)-\frac{Q^2}{Mr_0-Q^2}
	\end{equation}
which varies in the range $3<b<\infty$. The ISCO corresponds to $b=4$ and for stable orbits 
$3<b<4$. The two equations for $Z_1$ and $Z_3$ are coupled and possess one first integral 
generated by both the Killing vectors $K_t^{\alpha}$ and $K_z^{\alpha}$. On the 
other hand the equation for $Z_2$ is simple and may easily be integrated, what is actually 
unnecessary since its solutions are determined (without any integration) by the two first 
integrals generated by $K_x^{\alpha}$ and $K_y^{\alpha}$,
	\begin{equation}\label{n62}
Z_2(s)=C'\sin\left(\frac{L}{r_0^2}s\right)+C''\cos\left(\frac{L}{r_0^2}s\right)
	\end{equation}
and from (53) $L/r_0^2=\frac{1}{r_0}N^{-1/2}(Mr_0-Q^2)^{1/2}$.\\
The Jacobi vector field generated by $Z_2$ and vanishing at $P_0(s=0)$ is 
\begin{equation}\label{n63}
Z^{\mu}=C\delta^{\mu}_2\sin\left(\frac{L}{r_0^2}s\right)
\end{equation}
(this is the case 1 in sect.~2) and gives rise to the first infinite sequence of points 
$Q_n(s_n)$ conjugate to $P_0$ and located at distances 
\begin{equation}\label{n64}
s_n=n\pi\frac{r_0^2}{L}=n\pi r_0N^{1/2}(Mr_0-Q^2)^{-1/2}.
\end{equation}
The subsequent points $Q_n$ are separated by equal angular distance $\Delta\phi=\pi$. The 
deviation vectors spanned on the basis vectors $e_1^{\mu}$ and $e_3^{\mu}$ give rise to two 
other infinite sequences of conjugate points.\\
1. The second sequence (the case 2 in sect.~2) $Q'_n(s'_n)$ consists of points on B at 
distances (12) from $P_0$,
\begin{equation}\label{n65}
s'_n=2n\pi r_0^2N^{1/2}(MJ(r_0))^{-1/2},
\end{equation}
where $J(r)$ is defined as the left hand side of eq.~(57). To find out the spatial location 
of the first conjugate point $Q'_1$ one computes from (55) the ratio 
\begin{equation}\label{n66}
\frac{s'_1}{s_B}=r_0(Mr_0-Q^2)^{1/2}(MJ(r_0))^{-1/2};
\end{equation}
clearly $J(r_0)>0$ since $r_0>r_I$. For the orbits tending to the ISCO one has 
$r_0(Mr_0-Q^2)^{1/2}>r_I(Mr_I-M^2)^{1/2}>0$ and $J(r_0)\to 0$, then the twin B following a 
circular orbit close to the ISCO must make a large number of revolutions to encounter the 
first conjugate point $Q'_1$. For example, for $Q^2=0.5M^2$ and $r_0=5.17 M$ there is 
$s'_1/s_B=38.32$. The ratio quickly diminishes and again for $Q^2=0,5 M^2$ and $r_0=6M$ it 
is 2.760, nevertheless it is always larger than 1 and for $r_0\to \infty$ it tends to 1 from 
above. Qualitatively the behaviour is the same as in Schwarzschild spacetime \cite{S}.\\
2. For the third sequence (the case 3 in sect.~2) the eq.~(19) for the distance of $\bar{Q}_n$ 
from $P_0$ on B takes now on the form 
\begin{equation}\label{n67}
\bar{s}_n=r_0^2N^{1/2}(MJ(r_0))^{-1/2}\,y_n
\end{equation}
and from (65) one sees that the relative location of the conjugate points $Q'_n$ and 
$\bar{Q}_n$ is given by
\begin{equation}\label{n68}
\frac{\bar{s}_n}{s'_n}=\frac{y_n}{2n\pi}
\end{equation} 
and for $n\to\infty$ this ratio tends to $1+1/(2n)-4/(n^2\pi^2b)$ and for all $n$ it is larger 
than 1.

\section{Ultrastatic spherically symetric spacetimes}

The class of SSS spacetimes contains a special subclass of ultrastatic spherically symmetric 
(USSS) manifolds. We first provide a brief description of ultrastatic spacetimes. These are 
static spacetimes such that the unique timelike hypersurface orthogonal Killing vector field 
$K_t^{\alpha}$ is actually a gradient, $K_{t\alpha}=\partial_{\alpha}\Phi$. One chooses 
(in a non-unique way) a coordinate system $(x^{\alpha})$ in which the metric is 
time-independent, what is equivalent to $K_t^{\alpha}=\delta^{\alpha}_0$. Then 
$K_{t\alpha}=g_{\alpha0}=\partial_{\alpha}\Phi$ or $\Phi=Ct+f(x^k)$, with $C>0$ constant and 
an arbitrary function $f(x^k)$. Next one makes a coordinate transformation $t'=t+\frac{1}{C}
f(x^k)$ and $x'^i=x^i$ yielding $g'_{0i}=0$, $\Phi=Ct'$ and again $K'^{\alpha}_t=
\delta^{\alpha}_0$. Then rescaling time, $t'\to C^{-1/2}t$, and denoting the new 
coordinates by $x^{\alpha}$ one gets $ds^2=dt^2+g_{ij}(x^k)\,dx^i\,dx^j$ and the Killing vector has the constant length, $g_{\alpha\beta}K_t^{\alpha}K_t^{\beta}=+1$. 
Any USSS metric furthermore admits the three rotational Killing vectors, $K_x^{\alpha}$, 
$K_y^{\alpha}$ and $K_z^{\alpha}$ and in the standard chart it reads
	\begin{equation}\label{n69}
ds^2=dt^2-e^{\lambda(r)}dr^2-r^2\,d\Omega^2,
	\end{equation} 
in other words the metric function $\nu(r)=0$. These coordinates are at the same time the 
comoving ones, i.~e.~the metric in the comoving coordinates is time-independent. The 
best known USSS spacetime is the Einstein static universe with $\lambda=-\ln(1-r^2/a^2)$, 
where the constant $a$ has the dimension of length and $r\geq 0$; the spacetime 
physically is of merely historical significance since it is unstable. Other known 
USSS spacetimes do not include physically significant solutions, nevertheless they are 
interesting from the geometrical viewpoint because one is able to study not only radial 
and/or circular geodesic curves, but all properties of generic timelike geodesics 
including deviation vector fields and conjugate points. It is also interesting to notice 
that the comoving coordinates in each USSS spacetime imitate the inertial frame in 
special relativity: a free particle initially at rest remains always at rest and 
interparticle distances (between free particles at rest in the system) are 
time-independent. Furthermore, as we shall see below, a particle in a free motion has 
a constant velocity with respect to the comoving frame; the motion is not rectilinear 
since the notion is not well defined in these spacetimes. The similarity is not 
complete since there is a fundamental geometric difference: the space $t=$const in 
any USSS spacetime is not flat, although in the special cases where the space is the 
3--sphere $S^3$ or Lobatchevsky (hyperbolic) space $H^3$, it is homogeneous and 
isotropic (i.~e.~maximally symmetric). In other words the similarity of the system 
to the inertial frame concerns its dynamical properties.

A particle of mass $m$ moving on a timelike geodesic has the conserved energy $E$ and 
as previously $k\equiv E/(mc^2)$ and 
	\begin{equation}\label{n70}
\dot{t}\equiv \frac{dt}{ds}=k>0 \quad \Rightarrow \quad t-t_0=ks.
	\end{equation} 
The geodesic spatially lies in the 2--surface $\theta=\pi/2$, has the angular 
momentum $L$ and $\dot{\phi}\equiv d\phi/ds=L/r^2$. The radial component of the 
velocity $u^{\alpha}=dx^{\alpha}/ds$ satisfies 
	\begin{equation}\label{n71}
\ddot{r}+\frac{1}{2}\lambda'\dot{r}^2-\frac{L^2}{r^3}\,e^{-\lambda}=0
	\end{equation} 
($\lambda'=d\lambda/dr$) and from the integral of motion $u^{\alpha}u_{\alpha}=1$ is 
equal to 
	\begin{equation}\label{n72}
\dot{r}^2=e^{-\lambda}\left(k^2-1-\frac{L^2}{r^2}\right).
	\end{equation} 
The particle freely moving with respect to the coordinate frame has 3--velocity 
$v^i\equiv dx^i/dt$ with the constant modulus,
	\begin{equation}\label{n73}
|\mathbf{v}|^2=-g_{ij}v^iv^j=\frac{k^2-1}{k^2}<1,
	\end{equation} 
or the motion is uniform. 

From (72) one sees that $k^2-1\geq 0$. If $k=1 \Leftrightarrow E=mc^2$ one gets 
$\dot{r}^2\leq 0$ and this is possible only for $L=0$, the radial motion, which 
actually reduces to $\dot{r}=0$ and $r(s)=$const. The particle stands still in the 
space what means that no gravitational force is exerted on it; this surprising effect 
was first found in the special case of the global Barriola--Vilenkin monopole 
\cite{BV}, here we show that this is a common feature of all USSS spacetimes. 
Conversely, this fact implies that circular geodesics in any USSS manifold reduce 
to the trivial case of remaining at rest in the space. In fact, for $r=r_0=$const 
it follows from (71) that $L=0$, there is no azimuthal motion, then (72) implies 
$k=1$ and the geodesic is $t-t_0=s$, $r=r_0$, $\theta=\pi/2$ and $\phi=\phi_0$. \\
We therefore separately study radial and general non-radial geodesics.

\subsection{Radial geodesics}

By definition, the radial geodesic C has $L=0$ and is given by $t-t_0=ks$ with 
$k>1$, $\theta=\pi/2$ and $\phi=\phi_0$. We assume that the radial coordinate in 
the comoving system varies in the range $0\leq r_1\leq r\leq r_2\leq \infty$. 
Let the initial point of C be $P_0(t_0,r_0>r_1,\pi/2,\phi_0)$ with the initial 
velocity $\dot{r}^2(t_0)=u^2$. The formula (\ref{n72}) is reduced to 
	\begin{equation}\label{n74}
\dot{r}^2=(k^2-1)e^{-\lambda}.
	\end{equation} 
If $u>0$ then $u=+\sqrt{k^2-1}\,\exp(-\frac{1}{2}\lambda(r_0))$ and 
$\dot{r}=+\sqrt{k^2-1}\,\exp(-\frac{1}{2}\lambda(r))$ for $t>t_0$, similarly 
$\dot{r}^2=-\sqrt{k^2-1}\,\exp(-\frac{1}{2}\lambda(r))$ for $u<0$. In both the 
cases the sign of $\dot{r}$ remains unaltered for all times $t>t_0$ and the motion 
is unbounded, the radial coordinate either grows to the upper boundary $r_2$ or 
decreases to $r_1$ and this is accordance with the fact that the particle is subject 
to no gravitational force. From (\ref{n74}) one finds the length of the geodesic C between 
$r_0$ and $r(s)$,
	\begin{equation}\label{n75}
s(r)=\frac{\varepsilon}{\sqrt{k^2-1}}\,\int^r_{r_0}e^{\lambda/2}\,dr,
	\end{equation}
where $\varepsilon=+1$ for the outgoing geodesic and $\varepsilon=-1$ for the ingoing one. 
For some functions $\lambda(r)$ it is possible to invert this relationship to $r=r(s)$.

Now we seek for Jacobi fields on the radial C. Applying the general formalism \cite{SG1} we 
introduce the spacelike orthonormal basis triad which represents the special case $\nu=0$ 
of (\ref{n42}),
	\begin{equation}\label{n76}
e_1^{\mu}=[\varepsilon\sqrt{k^2-1},k,0,0], \qquad e_2^{\mu}=\frac{1}{r}\delta^{\mu}_2, \qquad
e_3^{\mu}=\frac{1}{r}\delta^{\mu}_3
	\end{equation}
and the equations for the Jacobi scalars are
	\begin{equation}\label{n77}
\frac{d^2Z_1}{ds^2}=0,
	\end{equation}
	\begin{equation}\label{n78}
\frac{d^2Z_2}{ds^2}=-\frac{k^2-1}{2r}\lambda'e^{-\lambda}\,Z_2
	\end{equation}
and the equation for $Z_3$ is identical to that for $Z_2$. The vector $Z^{\mu}=Z_1 e_1^{\mu}$ 
varies linearly since $Z_1=C_{11}s+C_{12}$. In (78) one replaces the derivative with respect to $s$ by derivatives w.~r~.~t.~ $r$ and denoting collectively $Z_2$ and $Z_3$ by $Z$ one gets 
	\begin{equation}\label{n79}
\frac{d^2Z}{dr^2}-\frac{1}{2}\lambda'\,\frac{dZ}{dr}+\frac{\lambda'}{2r}\,Z=0,
	\end{equation}
this equation may be solved for a given function $\lambda=\ln(-g_{11})$.

It is known from \cite{SG1} that the radial geodesics are globally maximal in the domain of the 
comoving coordinate chart, hence in the range $r_1<r<r_2$ they contain neither cut nor conjugate points to any of their points in this range.

\subsection{Properties of generic non--radial timelike geodesics}

A non--radial geodesic has $L\neq 0$ (we always assume that $L>0$). These curves require a separate 
treatment since most of their properties depend on $0<L^{-2}<\infty$. We denote such a geodesic by 
G (and as usual $\theta=\pi/2$ along it) and assume that it emanates from $P_0(t_0,r_0,\pi/2, \phi_0)$. 
Its behaviour depends on whether its initial radial velocity $\dot{r}(t_0)=u$ is positive or 
negative.\\
I. The initial velocity $u>0$.\\
Let $\lambda(r_0)=\lambda_0$. Then from (\ref{n72}) $u=+(k^2-1-L^2/r_0^2)^{1/2}\exp(-\lambda_0/2)$ 
and G is directed outwards and $r$ increases. Since $k^2-1-L/r^2>k^2-1-L/r_0^2>0$ for $r>r_0$, 
the radial velocity is always $\dot{r}>0$ (assuming $e^{-\lambda}>0$ for all finite $r$) and 
$r(s)$ is unbounded and tends to the chart boundary $r=r_2$. The non--radial G qualitatively 
behaves as the radial C. For $L\neq 0$ one introduces a parameter $p$ by $p^2\equiv (k^2-1)/L^2$, 
$0<p<\infty$. The length of G between $r_0$ and any $r\geq r_0$ is then    
\begin{equation}\label{n80}
s(r)=\frac{1}{L}\int^r_{r_0}\frac{r e^{\lambda/2}}{\sqrt{p^2r^2-1}}\,dr.
\end{equation}
In general this relationship cannot be inverted, therefore instead of $\phi(s)$ we seek for the 
dependence $\phi=\phi(r)$. From $d\phi/dr=\frac{L}{r^2}(dr/ds)^{-1}$ one easily derives 
\begin{equation}\label{n81}
\phi(r)-\phi_0=\int^r_{r_0}\frac{e^{\lambda/2}}{r}\frac{dr}{\sqrt{p^2r^2-1}}.
\end{equation}
II. The initial velocity $u<0$.\\
For some time interval after $t=t_0$ there is $\dot{r}=-(k^2-1-L^2/r^2)^{1/2}\exp(-\lambda/2)<0$ 
 and $r(s)$ monotonically decreases. Further evolution depends on the value of~$p$.\\
a) If $p^2\geq 1/r_1^2$ the geodesic G reaches (at least asymptotically) the lower boundary of the chart $r=r_1$ and its further evolution cannot be traced in the formalism we apply here. In this sense G is akin to the ingoing radial C.\\
b) If $p^2=1/r_m^2$ for some $r_m>r_1$ the geodesic G reaches the lowest point $r=r_m$ where its radial velocity vanishes, $\dot{r}(r_m)=-L\sqrt{p^2-1/r_m^2}\,\exp(-\lambda/2)=0$, 
while its azimuthal velocity $\dot{\phi}=L/r^2\neq 0$. The full 3--velocity squared is still 
equal to $(k^2-1)/k^2$ at $r_m$ and the particle is expelled outwards with positive radial 
acceleration following from (\ref{n71}),
\begin{displaymath}
\ddot{r}(r_m)=+\frac{L^2}{r_m^3}\,e^{-\lambda(r_m)}.
\end{displaymath} 
For later times $r(s)$ increases and correspondingly $\dot{r}>0$ increases and $\dot{\phi}$ 
diminishes. The geodesic G qualitatively behaves as that emanating from $P_0$ with $u>0$. 
Analogously to (\ref{n80}) one finds that the length of G from $r_0$ down to $r_m$ and then upwards 
to $r$ is 
\begin{equation}\label{n82}
s(r_0,r_m,r)=\frac{1}{L}\left(\int^{r_0}_{r_m}+\int^r_{r_m}\right)\frac{r\,e^{\lambda/2}}
{\sqrt{p^2r^2-1}}\,dr
\end{equation}
and if the indefinite integral of the integrand is $F(r)$, then $s(r_0,r_m,r)=F(r)-F(r_0)-
2F(r_m)$. The $\phi(r)$ dependence may be found in the similar way.

\subsection{The geodesic deviation equations for a generic non--radial geodesic}

For the sake of simplicity we consider geodesics consisting only of one segment, i.~e.~the 
curves emanating outwards ($\dot{r}>0$) from the initial $P_0$, then $t-t_0=ks$ and (\ref{n80}) and (\ref{n81}) hold. The vector tangent to G is
	\begin{equation}\label{n83}
u^{\alpha}=[k,\frac{L}{r}e^{-\lambda/2}(p^2r^2-1)^{1/2},0,\frac{L}{r^2}].
	\end{equation} 
(Here and below always $p^2r^2>1$.) It is not easy to find the spacelike basis triad 
possessing the required properties on G \cite{SG1} and only after many manipulations and solving 
a system of two coupled first order ODEs one arrives at
	\begin{eqnarray}\label{n84}
e_1^{\mu} & = & \left[0,\frac{1}{pr}e^{-\lambda/2}, 0, -\frac{1}{pr^2}\sqrt{p^2r^2-1}\right], \qquad 
e_2^{\mu}=\frac{1}{r}\delta^{\mu}_2,\nonumber\\
e_3^{\mu} & = & \left[\sqrt{k^2-1}, \frac{k}{pr}e^{-\lambda/2}\sqrt{p^2r^2-1}, 0, \frac{k}{pr^2}\right].
	\end{eqnarray} 
The curvature tensor for the metric (\ref{n69}) is 
	\begin{equation}\label{n85}
R_{1212}=-\frac{r}{2}\lambda', \quad R_{1313}=-\frac{r}{2}\lambda'\sin^2\theta, \quad 
R_{2323}=r^2(e^{-\lambda}-1)\sin^2\theta.
	\end{equation}
Applying (\ref{n83}), (\ref{n84}) and (\ref{n85}) one derives the equations for the Jacobi scalars,
\begin{equation}\label{n86}
\frac{d^2Z_1}{ds^2}=-\frac{1}{2}(k^2-1)\frac{\lambda'}{r}e^{-\lambda}\,Z_1,
	\end{equation}
	\begin{equation}\label{n87}
\frac{d^2Z_2}{ds^2}=\frac{L^2}{r^3}\left[\frac{1}{r}(e^{-\lambda}-1)-\frac{\lambda'}{2}e^{-\lambda}
(p^2r^2-1)\right]\,Z_2,
	\end{equation}
	\begin{equation}\label{n88}
\frac{d^2Z_3}{ds^2}=0.
	\end{equation}
Since $Z_3=C_{31}s+C_{32}$ one sees that the deviation vector $Z^{\mu}=Z_3\,e_3^{\mu}$ does not 
generate conjugate points on G. To make the equations for $Z_1$ and $Z_2$ solvable one 
replaces $d/ds$ by $d/dr$, then the equivalent equations read
\begin{equation}\label{n89}
\frac{d^2Z_1}{dr^2}+\left[
\frac{1}{r(p^2 r^2-1)}-\frac{\lambda'}{2}\right]\frac{d Z_1}{dr}+\frac{1}{2}\frac{p^2 r\lambda'}{p^2 r^2-1}Z_1=0,
\end{equation}
\begin{equation}\label{n90}
\frac{d^2Z_2}{dr^2}+\left[
\frac{1}{r(p^2 r^2-1)}-\frac{\lambda'}{2}\right]\frac{d Z_2}{dr}-\frac{1}{r}
\left[\frac{1-e^\lambda}{r(p^2 r^2-1)}-\frac{\lambda'}{2}\right]Z_2=0.
\end{equation}
The timelike and the three rotational Killing vector fields generate four integral of 
motion; that generated by $K_t^{\alpha}=\delta^{\alpha}_0$ gives rise to the solution for $Z_3$. 
The vectors $K_x^{\alpha}$ and $K_y^{\alpha}$ generate two independent first integrals for 
eq.~(\ref{n90}),
	\begin{equation}\label{n91}
f(r)\frac{dZ_2}{dr}\,\sin\phi-\frac{1}{r}\left(f(r)\sin\phi+\cos\phi\right)Z_2=C_1,
	\end{equation}
\begin{equation}\label{n92}
-f(r)\frac{dZ_2}{dr}\,\cos\phi+\frac{1}{r}\left(f(r)\cos\phi-\sin\phi\right)Z_2=C_2
\end{equation}
and multiplying (\ref{n91}) by $\cos\phi$ and (\ref{n92}) by $\sin\phi$ and adding the two equations one gets $Z_2$ without any integration; the general solution is 
\begin{equation}\label{n93}
Z_2(r)=C_{21}\,r \cos\phi(r) +C_{22}\, r\sin\phi(r)
\end{equation}
here $\phi(r)$ is given by (\ref{n81}). A first integral for $Z_1$ is generated by $K_z^{\alpha}=
\delta^{\alpha}_3$ and reads
\begin{equation}\label{n94}
\frac{1}{r}(p^2r^2-1)\,\frac{dZ_1}{dr}-p^2\,Z_1=C_1e^{\lambda/2}
\end{equation}
and is easily solved by the following general solution
\begin{equation}\label{n95}
Z_1(r)=C_{11}\sqrt{p^2 r^2-1}+C_{12}\sqrt{p^2 r^2-1}\int r e^{\lambda/2}(p^2 r^2-1)^{-3/2}dr.
\end{equation}
Without knowledge of $\lambda(r)$ one can say very little about this solution. Yet concerning 
$Z_2(r)$ one can find a generic condition for the existence of conjugate points to $P_0$ 
generated by $Z^{\mu}=Z_2e_2^{\mu}=(C_{21}\cos\phi+C_{22}\sin\phi)\delta^{\mu}_2$. This vector 
is directed off the 2--surface $\theta=\pi/2$. Without loss of generality one can rotate the 
coordinate system in the surface to put $\phi_0=0$. Then the deviation vector vanishing for 
$\phi=0$ is 
\begin{equation}\label{n96}
Z^{\mu}(r)=C_{22}\delta_2^\mu\sin\phi(r)
\end{equation}
and is zero at points $\phi_n(r)=n\pi$, $n=1,2, \ldots$. The first conjugate point exists iff the 
difference $\phi(r)-\phi_0$ given in (81) exceeds $\pi$ for some $r<\infty$ and this may be 
concluded only when $\lambda(r)$ is known.\\
In this work we discuss only one special case of USSS spacetimes and postpone other examples to a 
forthcoming paper.

\section{The global Barriola--Vilenkin monopole}

This is an approximate solution of Einstein's field equations with a source being a triplet of 
scalar fields, describing the spacetime outside a monopole resulting from the breaking of a 
global $O(3)$ symmetry of the triplet \cite{BV}. Assuming that in the astrophysically relevant 
situations the monopole mass is negligibly small, the resulting metric is approximated by \cite{BV} 
	\begin{equation}\label{n97}
ds^2=dt^2-\frac{1}{h^2}dr^2-r^2 d\Omega^2,
	\end{equation}
here $r$ is larger than some $r_1>0$ and $h$ is a dimensionless constant, $0<h<1$ (to fit the general USSS formalism of sect.~4 we have rescaled the radial coordinate applied in \cite{BV}). 
The metric function $\lambda=-2\ln h=$const.\\
The general outgoing ($\dot{r}>0$) non--radial geodesic has the length 
	\begin{equation}\label{n98}
s(r)=\frac{1}{h p^2 L}\left(
\sqrt{p^2 r^2-1}-\sqrt{p^2 r_0^2-1}
\right)
	\end{equation} 
and this function may be inverted to 
\begin{equation}\label{n99}
r^2(s)=(h p L)^2 s^2+2 h L\sqrt{p^2r_0^2-1}\,s+r_0^2
\end{equation}
whereas the integral (\ref{n81}) is 
\begin{equation}\label{n100}
\phi(r)-\phi_0=\frac{1}{h}\left[\arccos\left[\frac{1}{pr}
\right] - \arccos\left[\frac{1}{p r_0} \right]
\right]
\end{equation}
here $0<1/(pr)\leq 1/(pr_0)\leq 1$. That the spacetime admits no bound orbits was found by 
Chakraborty \cite{Ch} using the Hamilton--Jacobi formalism; the coordinates of a timelike 
geodesic were given as integrals with respect to the radial coordinate without employing the 
fact that the motion is `flat'.\\
1. The outgoing radial geodesic C has the length 
\begin{equation}\label{n101}
s(r)=\frac{r-r_0}{h\sqrt{k^2-1}}\quad \mbox{and}\quad r-r_0=h\sqrt{k^2-1}\,s
\end{equation} 
For $\lambda'=0$ the equations for the Jacobi scalars reduce to $d^2Z_a/ds^2=0$, hence 
$Z_a(s)=C_{a1}s+C_{a2}$ and the distances between the radial geodesics (all in $\theta=\pi/2$) 
grow linearly.\\
2. For the non--radial ($L>0$) outgoing geodesic G the Jacobi scalars are
\begin{equation}\label{n102}
Z_1(r)=C_{11}\sqrt{p^2r^2-1}+C_{12}
\end{equation}
from (95), $Z_2$ is as in (93) and $Z_3=C_{31}s+C_{32}$. We check whether $Z_1$ generates a 
conjugate point. We put $Z_2=Z_3=0$ and require $Z_1(r_0)=0$, this yields (for $C_{11}=1$) 
	\begin{equation}\label{n103}
Z_1(r)=\sqrt{p^2r^2-1}-\sqrt{p^2r_0^2-1}
	\end{equation}
and this function has no roots for $r>r_0$. It remains to study the vector $Z^{\mu}=Z_2
e_2^{\mu}$ and we know that the candidates for the conjugate points are at $\phi_n=n\pi$. 
We denote $\alpha\equiv \arccos(1/(pr_0))$ and since $pr_0\geq 1$ one gets $0<\alpha\geq 
\pi/2$ and setting $\phi_0=0$ eq.~(100) is replaced by 
	\begin{equation}\label{n104}
\phi(r)=\frac{1}{h}\left[\arccos\left[\frac{1}{pr}\right]-\alpha\right].
	\end{equation}
For $r\to\infty$ this expression tends to $\frac{1}{h}(\pi/2-\alpha)$ and for $\alpha<\pi/2$ and 
$h$ sufficiently small it is arbitrarily large admitting many conjugate points. The first 
conjugate point $Q_1$ to $P_0$ occurs for $r=r_c$ such that \\
$\arccos(\frac{1}{pr_c})-\alpha=h\pi$\\
and the obvious inequality $0<\arccos(\frac{1}{pr_c})<\pi/2$ is equivalent to $0<h\pi+\alpha<
\pi/2$ or $0<\alpha<\pi/2-h\pi$. The last inequality means that $\cos\alpha=1/(pr_0)>\cos(\pi/2-
h\pi)=\sin h\pi$. The value of $h$ is fixed and assigned to the spacetime. Therefore the 
necessary condition for the existence of the conjugate points on some timelike geodesics is 
$h<1/2$. If it holds then the sufficient condition for the geodesic G to contain the conjugate 
point $Q_1$ (and possibly the successive points) to $P_0$ is 
\begin{equation}\label{n105}
pr_0<\frac{1}{\sin h\pi}.
\end{equation}
It is worth stressing that all these properties hold if the spacetime outside the monopole is 
described in the astrophysical approximation of negligible mass. If $M\neq 0$ all the 
geometrical features of the resulting spacetime are significantly altered. For a positive mass 
one gets an SSS manifold outside the USSS subclass and its metric may be rescaled to take on the 
Schwarzschild form implyinjg the geodesic structure as in the latter spacetime, whereas for 
$M<0$ the gravitation is repulsive and there are no bound orbits \cite{HL}.

\section{Conclusions}
Our conclusions to this work are brief and partial because any general rules concerning the 
qualitative properties of timelike worldlines in various spacetimes may be formulated only 
after completing the research programme, i.~e.~after studying a sufficiently large number of 
different Lorentzian manifolds. In our opinion this statement should not be interpreted as a 
declaration that the research programme will never terminate. On the contrary, we believe that 
the wealth of geodesic stuctures in various gravitational fields is not immense and can be 
grasped in its dominant features in a not far future.\\
We have shown in the previous three papers that static spherically symmetric spacetimes are 
different in some properties of timelike geodesic curves and very similar in many other ones. 
Here we show that for all stable circular orbits in these spacetimes all the infinitesimally 
close geodesics to the circular ones, constructed by means of the geodesic deviation vector 
fields, have the same length between a pair of conjugate points on the circular curves. The 
radial and circular geodesics in Reissner--Nordstr\"{o}m spacetime behave similarly to those 
in the Schwarzschild field, in particular if a circular and radial geodesics have common 
endpoints, the radial is longer. Spherically symmetric spacetimes which furthermore are 
ultrastatic, exhibit a rather astonishing property that a free test particle may be in the 
state of absolute rest, i.~e.~no gravitational force is exerted on it. If a free particle 
moves, its velocity is constant---this means that the comoving frame in these spacetimes is 
dynamically akin to the inertial frame in Minkowski space of special relativity. Here we 
have studied one example of ultrastatic manifolds, the Barriola--Vilenkin global (massless) 
monopole; the metric contains one arbitrary function which in this case is a constant and a 
generic non-radial geodesic may have conjugate points if the constant is sufficiently small.

\textbf{Acknowledgements}\\
We are grateful to our colleagues from the Krakow Relativity Group for valuable comments. This 
work was supported by a grant from the John Templeton Foundation.

\end{document}